# Magnetic ordering in the fine-particles of some bulk Pauli-paramagnets

Sitikantha D. Das, S. Narayana Jammalamadaka, Kartik K Iyer, and E.V. Sampathkumaran[*]

Tata Institute of Fundamental Research, Homi Bhabha Road, Mumbai 400005, India

Fine particles of some of the extensively studied 'enhanced' Pauli-paramagnets - $LuCo_2$, $ScCo_2$, and $ZrCo_2$ and $CeRu_2Si_2$, have been prepared by high-energy ball-milling and investigated for their magnetic behavior. In the case of Co systems, these particles show ferromagnetic features even at room temperature. Even in the case of $CeRu_2Si_2$, a well-known Kondo lattice which is nonmagnetic in the bulk form, features attributable to the onset of magnetic order below about 8 K could be seen in fine particles. The magnitudes of the magnetic moment of these small particles at high fields are sufficiently large to conclude that the observed magnetic ordering is not just a surface effect. This transformation to magnetic ordering in small particles could be widespread among enhanced Pauli paramagnets.

PACS numbers: 73.63.Bd; 75.30.Kz; 75.50.Tt; 71.20.Eh

# I. INTRODUCTION

Despite intense research in the fields of nanomagnetism and strongly correlated electron systems (SCES), there appears to be very little effort to understand electron correlation behavior in nanoparticles of SCES, barring a few recent reports [1, 2]. It was found that the disorder in fine-particles of SCES systems studied till todate [1] tends to destabilize magnetic ordering. It is interesting that this is true even in Fe, Co and Ni which are found to become nonmagnetic in nanoform [3]. In light of such recent developments, it is important to address this question in other magnetic systems to evolve a global scenario.

In this article, we report the results of our systematic investigations [4] on the fine particles of some of the most extensively studied 'enhanced' Pauli-paramagnets – three of them belonging to the Laves-phase family [5] viz., $LuCo_2$, $ScCo_2$, and $ZrCo_2$ [6,7] and one of great current interest [8, 9] belonging to heavy-fermion family, $CeRu_2Si_2$, crystallizing in $ThCr_2Si_2$-type tetragonal structure. These Lu [6], Zr [7] and Ce [10, 11] systems have been known to undergo itinerant-electron metamagnetism (IEM), that is, crossover to ferromagnetism by the application of magnetic fields ($H$) around 800, 600 and 75 kOe respectively; however, the critical field for the Sc compound has been predicted to be more than 1200 kOe [12], though, to our knowledge, there is no experimental verification of the same till to date. We report here that, in the fine-particle form, the ferromagnetism above 300 K in all these Co-based enhanced paramagnets *including Sc case* and some kind of magnetic ordering near 8 K for the Ce system could be induced. The results bring out that a deviation from the trend briefed in the introductory paragraph could be specific to enhanced Pauli-paramagnets.

# II. EXPERIMENTAL DETAILS

The polycrystalline ingots of $LuCo_2$, $ScCo_2$, $ZrCo_2$ and $CeRu_2Si_2$ were prepared by repeated melting of required amounts of high purity (>99.9%) constituent elements in an arc furnace in an atmosphere of argon. The weight losses after final melting were negligible (<0.5%). The specimens thus obtained were characterized by x-ray diffraction (XRD) (Cu $K_\alpha$) to be single phase (within the detection limit of 2%), except for $ScCo_2$, in which case a weak extra line appears as shown in figure 1 by an asterisk. We performed magnetization (*M*) studies on these ingots as a function of temperature (*T)* and *H* to make sure that the starting materials are Pauli-paramagnets and there is no interference from any ferromagnetic impurity. The parent ingots (called *A*) thus synthesized were used to prepare the small particles. The materials were milled in a planetary ball-mill (Fritsch pulverisette-7 premium line) operating at a speed of 500 rpm in a medium of toluene for different intervals of time and corresponding specimens are labeled *B*, *C*, and *D*. Unless otherwise stated, tungsten carbide (W-C) vials and balls of 5 mm diameter were used with a balls-to-material mass ratio of 10:1. We monitored phase-purity by XRD patterns. The scanning electron microscopic (SEM, Nova NanoSEM600, FEI Company) pictures confirmed homogeneity of all the single-phase samples and energy dispersive x-ray analysis (EDXA) confirmed that the atomic ratios of the constituent elements are the same as those of ingots. The same SEM and a transmission electron microscope (TEM, Tecnai 200 kV) were employed to confirm the reduction of particles to nanometer size. The *dc M(T)* (in the range 1.8-300 K) and *M(H)* (up to 50 kOe) measurements at selected temperatures for all specimens were carried out with the help of a commercial superconducting quantum interference device (Quantum Design) and the same magnetometer was employed to take *ac* susceptibility *(χ)* data for $CeRu_2Si_2$. In addition, for the Ce sample, we have employed a commercial vibrating sample magnetometer (Oxford Instruments) as there is a need to extend *M(H)* measurements at 1.8 K to higher fields.

### III. RESULTS AND DISCUSSION

The XRD patterns are shown in figure 1 for different milling times, along with respective patterns for the molten ingot. A general feature is that the lines are distinctly sharp and there is no evidence for any change in the shapes of the background curves, thereby implying that the specimens remain crystalline in all cases. However, there is a reduction in intensity with increasing milling time as reported in the literature commonly for ball-milled metals. The region around the most intense line is shown (but not for all specimens for the sake of clarity) in the expanded form in the insets of the figure 1 after normalizing to peak heights to highlight the fact that the lines tend to broaden with increasing milling time. It is noteworthy that there is no change in the lattice constants within the limits of experimental error ($\pm$ 0.004 Å) with increasing milling time. The x-ray diffraction patterns of the powders exposed to air for over a long period of time (say, for a few months) indicate that the specimens do not deteriorate as long as the toluene layer is not removed, say, by heating at 100 C. For $ScCo_2$, an extra line that is present in the ingot disappeared after milling for 5 minutes resulting in a single phase. For specimen *D* of $ScCo_2$ and $ZrCo_2$ (see curve *D* in figure 1 for these compounds), another phase appears, which is weak (<5%) for the former. Otherwise, in the rest of the ball-milled specimens, there is no evidence for free Co or any other magnetic phase and therefore the magnetic behavior reported below is characteristic of fine-particles of respective compounds. For such single-phase specimens, we have performed Reitveld analysis to ensure proper stoichiometry and site occupation and the refinement parameters fall in a satisfactory range ($R_{wp}$ ~ 8; $R_p$ ~ 6; $\chi^2$ ~ 1 with these parameters implying usual meanings in such an analysis). An idea of the average particle sizes could be inferred from the width of the most intense line (after subtracting instrumental line-broadening) employing Debye-Scherrer formula, and the values thus obtained typically are 100-110 nm for *D* of $LuCo_2$ and $ScCo_2$ and 20-30 nm for *D* of $ZrCo_2$ and $CeRu_2Si_2$. Attempts to correct for strain effects employing Williamson-Hall plots were not successful, as such plots are not found to be linear, possibly due to a large spread in the particle sizes. Therefore, the values should be treated with some caution and serve as a qualitative guide only. In order to establish that one attains particles in the nanometer range, we show representative SEM and bright-field TEM images, for instance, for $LuCo_2$ and $CeRu_2Si_2$ in figure 2. SEM images reveal significant agglomeration of the particles. We have isolated some of the particles by ultrasonification in alcohol and the TEM images confirm the formation of the particles with (irregular) sizes in the nanometer range. For the purpose of this article, the actual size of the particles or its distribution is irrelevant.

*We now discuss the magnetization results on the Co-based systems, $LuCo_2$, $ScCo_2$, and $ZrCo_2$*:

The results of magnetization measurements in the temperature range 4.2 – 300K are shown in Fig. 3 for all Co specimens. In the molten ingots of $LuCo_2$ and $ScCo_2$, if one magnifies (not shown here) the plot of *M/H*, it increases with increasing temperature as though there is a maximum at higher temperatures, typical of exchange-enhanced paramagnets in agreement with the literature (See Ikeda et al, Ref. 5). In the case of the ingot of $ZrCo_2$, as in Ref. 7, *M/H* is found to increase with decreasing temperature with a broad maximum around 40 K. The magnitudes of *M/H* increase as one traverses from specimens *A* to *D* in all cases and it is to be noted that the sign of *dM/dT* changes (to negative) for the milled samples of $LuCo_2$ and $ScCo_2$. The values of *M/H* for the specimens milled even for shorter durations (e.g., 5 to 30 mins) are found to be significantly larger when compared with those of the respective parent ingots. *M/H* gradually decreases with increasing *T* in all fine-particles which could be a signal for ferromagnetic order above 300 K. [In fact, this could be visually verified by the fact that the ball-milled specimens get attracted to a small bar magnet at room temperature]. In order to understand the magnetic behavior better, in figure 3 (right) as well as in figure 4, we show the *M(H)* curves at 300 K and 4.2 K. There is a steep increase in the magnetization for all the specimens for an initial application of magnetic fields and *M(H)* curves are found to be hysteretic at low-fields consistent with ferromagnetism. At higher fields, there is a relatively much weaker variation with *H* without saturation. Absence of complete saturation could imply that there is a superparamagnetic component as well. The values of *M* extrapolated from the high-field data to zero-field gets gradually more significant with increasing milling-time, as though ferromagnetic component tends to dominate gradually. The values of extrapolated saturation moment ($\mu_s$) at 4.2 K for final single-phase specimens fall in the range 0.4 - 0.8 $\mu_B$/formula-unit, which are comparable to those known in the literature for the bulk forms after field-induced metamagnetic transitions [6]. From these large magnitudes, it is evident that the observed magnetism is neither surface effect nor due to traces of magnetic impurity, as otherwise the magnitudes would be two orders of magnitude smaller [13, 14]. It should be noted that there are even recent claims that the magnetization should get suppressed in the surface of nanomaterials [15].

From the hysteresis loops shown in figure 3 (right), it is obvious that coercive field ($H_c$) is less than 100 Oe at 4.2 K. $H_c$ increases with decreasing temperature. The value is also found to decrease with increasing milling time (not shown here). This trend signals that the size of the particles is already in the single-domain region [16].

In short, the magnetism of all these fine-particles can be described in terms of a mixture of superparamagnetic and ferromagnetic behavior with a Curie temperature above 300 K.

*We now discuss the behavior of fine-particles of $CeRu_2Si_2$.*

From the XRD patterns, we infer that the milled-specimens (even for 6 hours milling) remain crystalline. In fact, we did not see any significant change in the diffraction-line intensities as the milling-time is increased from one and half hour (see figure 1) to 6 hours (not shown here). The line-widths for *C* and *D* also remain nearly the same, as though milling well beyond 30 minutes does not significantly change the particle size in this compound. Thus, it is inferred that one could obtain fine-particles within 90 minutes of milling.

With respect to the magnetic behavior of the fine-particles (see figure 5), there is a sudden change in the slope of $d\chi/dT$ in the *dc* $\chi$ data, around 8 K for all milled specimens. $\chi(T)$ curves obtained in a field of 100 Oe for the zero-field-cooled (*ZFC,* from 35 K) and field-cooled (FC) conditions of the specimens tend to bifurcate at this temperature. Interestingly, the values of $\chi$ and the features are found to be nearly the same for all specimens milled during 30 minutes to 6 hour and for this reason we present the data for specimen *D* only. In ZFC curve, there are two peaks, one near 6 K and the other at 2.5 K, which are modified in FC curve, as though the magnetic state is a complex one. There are corresponding anomalies in the *ac χ(T)* data as well ($H_{ac}$= 1 Oe) as shown in figure 5 for both real and imaginary parts. We could not resolve any frequency dependence of the peak temperatures, as the peaks are rather broad, possibly due to inhomogeneous magnetism arising from particle-size distribution. Therefore, though these results establish the existence of a magnetic transition near 8 K, we could not confirm from these data whether the ordering is of a ferromagnetic type or a spin-glass type. We have therefore measured isothermal remanent magnetization, $M_{IRM}$, at 1.8 and 5 K. For this purpose, the specimens were zero-field-cooled to the desired temperature, a magnetic field (say, 10 kOe) was switched on for 5 mins and then $M_{IRM}$ was measured after the field was switched off. We note that $M_{IRM}$ decays slowly with time (logarithmically after initial few minutes) as shown in an inset of figure 5. While this slow decay is qualitatively consistent with spin-glass freezing, it is at present not clear why the decay is faster at 5.5 K compared to that at 1.8 K. This may indicate complex nature of magnetism, suggesting a subtle change in magnetism across 2.5 K We have also measured isothermal *M* behavior at 1.8 K up to 120 kOe to throw more light on magnetism (see figure 4). *M(H)* behavior of the bulk form is in conformity with that known for polycrystals in the literature [10] in the sense that there is a broad feature attributable to spin reorientation around 70 kOe. However, for the milled-specimens, the nature of *M(H)* curve is completely modified. An upward curvature appears for all specimens and the feature attributable to spin-reorientation disappears. In addition, the curve is hysteretic, though weak. Thus, a ferromagnetic component seems to develop in the fine-particles. However, there is no evidence for saturation of M till 120 kOe. These observations viewed together favor a spin-glass-type of ordering, though exact nature of magnetic structure has to be ascertained by neutron diffraction studies. We also note that, to attain this magnetic behavior, it does not matter whether the milling is done for 30 minutes or for an extended period, as *M(H)* curves nearly overlap for all milled-specimens (and hence not shown for all specimens for the sake of clarity). The value of the magnetic-moment obtained at high fields, say at 120 kOe (about 0.5 $\mu_B$/Ce) is significant enough to conclude that the observed features and inferences can not be attributable to surface or impurity. Finally, it is worth noting that a lattice expansion caused by a small replacement of Si by Ge or Ce by La in $CeRu_2Si_2$ results in a magnetic transition [11] in the same temperature range as that in the fine-particles. This comparison implies that a reduction in size corresponds to lattice expansion.

The low-field hysteresis data (± 10 kOe) at 300 K on the above-mentioned milled specimens reveal the presence of a strong paramagnetic component superimposed over a weak ferromagnetic component. The value of the saturation moment extrapolated to zero field from high fields turns out to be <0.002 $\mu_B$/formula-unit. [This was found to be true even for the fine particles of the La analogue]. We suspected that a small ferromagnetic impurity was possibly introduced by Co present in the W-C vials. Therefore, we prepared additional specimens with zirconia vials and balls and found that this very weak ferromagnetic component is absent in such specimens. The plot of inverse χ versus T for such a specimen overlaps with that of the molten ingot as shown in the inset of figure 5 and found to obey Curie-Weiss law above ~125 K. The value of the effective moment (~2.4 $\mu_B$/Ce) obtained from the Curie-Weiss region turns out to be very close to that expected for trivalent Ce ion. The paramagnetic Curie temperature is about 38 K as in our bulk sample. As the temperature is lowered below 10 K, the features in *ac* and *dc* χ were found to be the same as those obtained with WC vial.

## IV. SUMMARY

We have demonstrated that magnetic order in fine-particles is widespread among those compounds which are 'enhanced' Pauli-paramagnets in bulk form. It is possible that the observed magnetic transition in the fine-particles of the present materials is controlled by electronic structure [7]. In fact, the narrowing of the d-band due to defects and subsequent shift of the d states to Fermi energy has been proposed to play a crucial role favoring itinerant electron magnetism in the chemically-doped $RCo_2$ [17]. It is possible that such factors under favorable conditions stabilize, rather than destabilizing, magnetic ordering in fine-particles prepared by ball-milling. This study in general opens up a new route for understanding of SCES in nanoforms by many other techniques.

## ACKNOWLEDGEMENTS

The authors thank N.R. Selvi Jawaharlal Nehru Center for Advanced Scientific Research, Bangalore, India, for SEM data and B.A. Chalke and S.C. Purandare for characterization of the specimens by TEM studies.

**Note:**
We have taken selected area electron diffraction patterns employing TEM on some milled specimens, for instance, for $ZrCo_2$ and $CeRu_2Si_2$ compounds (see figures below). The diffraction rings could be indexable to respective crystallographic phases and there was no evidence for any other phase. This endorses that the properties presented in this paper for the fine particles correspond to the compounds under discussion.

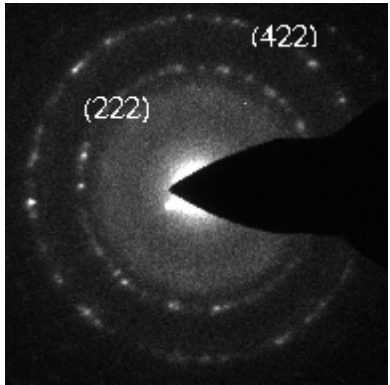
$ZrCo_2$

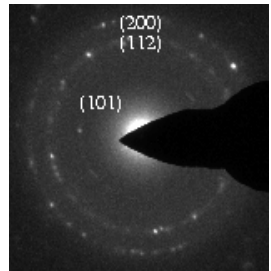
$CeRu_2Si_2$

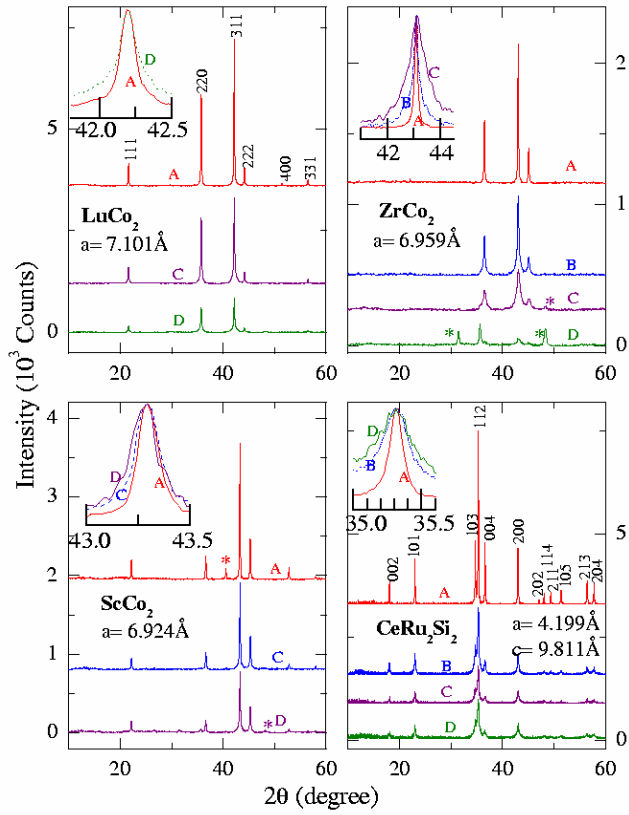

Figure 1:
(color online) X-ray diffraction patterns of the molten ingots and some of the ball-milled specimens of LuCo$_2$, ScCo$_2$, ZrCo$_2$ and CeRu$_2$Si$_2$. In the insets, for each compound, the most intense peaks normalized to the peak-heights are plotted for some specimens to show how lines broaden after milling. The notations *A, B, C,* and *D* are marked for specimens with increasing milling time; *A* corresponds to molten ingot in all cases; while *C* and *D* for LuCo$_2$ and ScCo$_2$ correspond to those specimens milled for 20 and 45 minutes respectively, *B*, *C* and *D* specimens for ZrCo$_2$ were obtained by milling for 30, 90 and 270 minutes; for CeRu$_2$Si$_2$, *B*, *C* and *D* correspond to 30, 90 and 150 minutes milling. The lattice constants (a, c, ±0.004 Å) are also included.

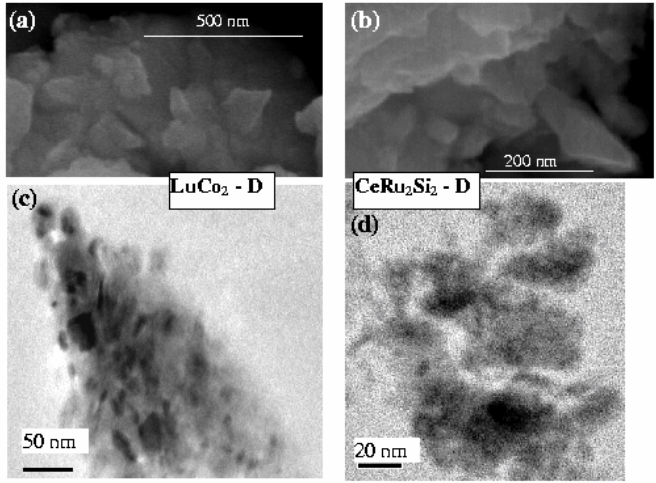

Figure 2:
Scanning electron microscopic pictures (**a** and **b**) and bright-field images (**c** and **d**) of transmission electron microscope obtained on the ball-milled specimens *D* of $LuCo_2$ and $CeRu_2Si_2$.

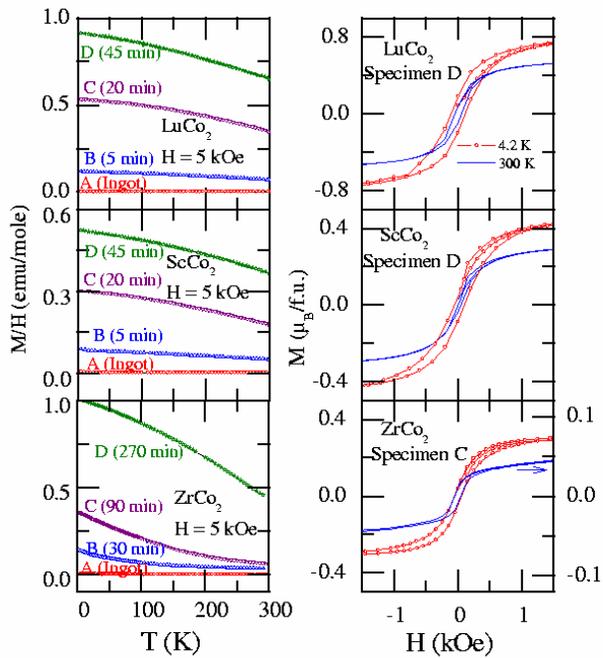

Figure 3:
(color online) (Left) The values of *dc* magnetization *(M)* divided by magnetic-field *(H)* are plotted as a function of temperature for the molten ingots and ball-milled specimens of $LuCo_2$, $ScCo_2$, and $ZrCo_2$. (Right) Corresponding low-field hysteresis loops at 4.2 and 300 K.

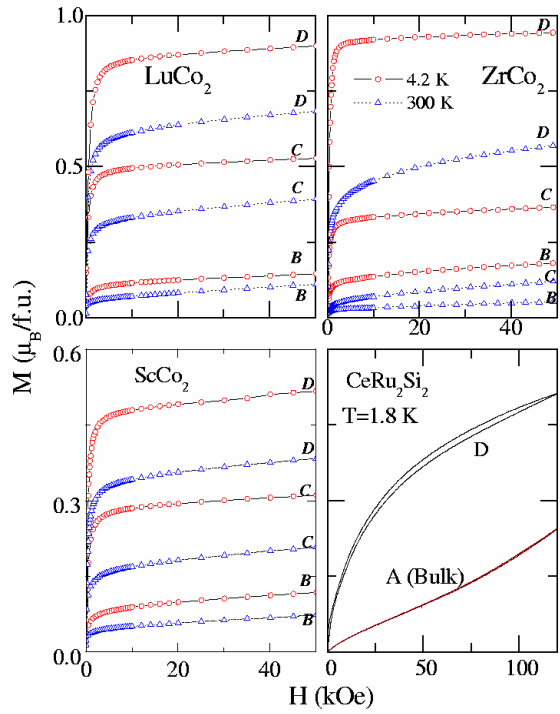

Figure 4:
(color online) Isothermal magnetization per formula unit at 4.2 and 300 K for the ingots and ball-milled specimens of LuCo$_2$, ScCo$_2$, ZrCo$_2$ and CeRu$_2$Si$_2$.

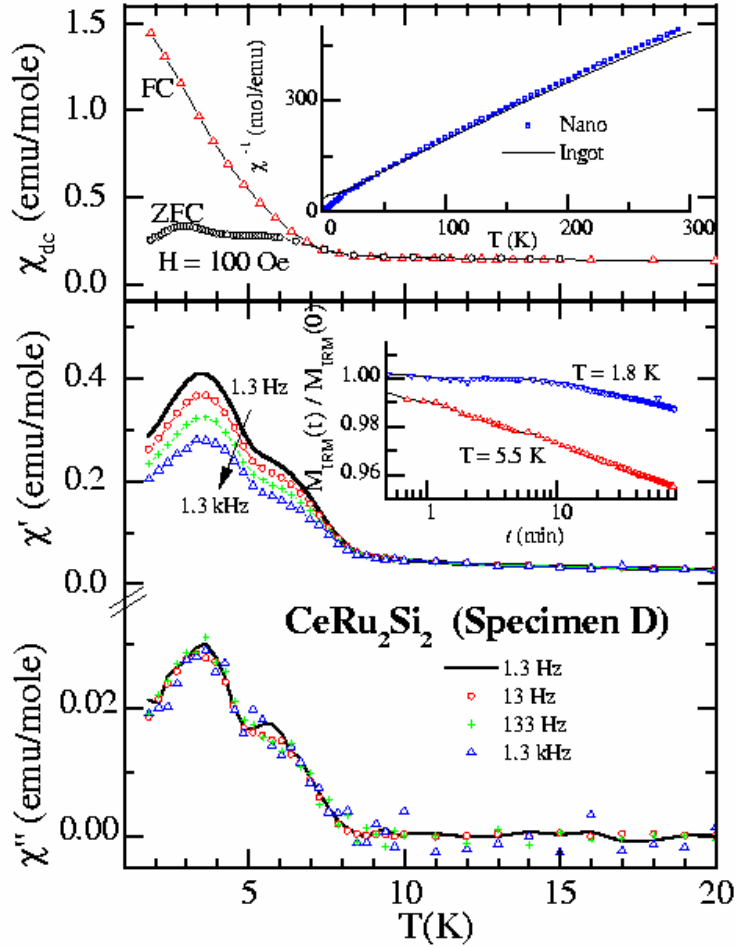

Figure 5:
(color online) In the mainframe, magnetic susceptibility ($\chi$) as a function of temperature measured in the presence of 100 Oe for the zero-field-cooled (*ZFC*) and field-cooled (*FC*) conditions of the sample, and real ($\chi'$) and imaginary ($\chi''$) parts of *ac* $\chi$, for the milled-specimen *D* of $CeRu_2Si_2$. Inverse *dc* susceptibility as a function of temperature (for *H*= 5 kOe) for the molten ingot as well as for the specimen obtained by ball-milling using a zirconia vial and normalized time-dependent isothermal remnant magnetization for specimen *D* are shown in the insets.